\newif\ifsubmode
\submodetrue
 
\ifsubmode
  \documentclass[12pt,preprint]{aastex}  
  \received{}
  \revised{}
  \accepted{}
 
\else
  \documentclass{emulateapj}  
  \slugcomment{Submitted to \apjl}
\fi

\def\vol#1  {{{#1}{\rm,}\ }}

\def\apj    {{ApJ{\rm,}\ }}
\def\apjl   {{ApJ{\rm,}\ }}

\def\aj     {{AJ{\rm,}\ }}
\def\aa     {{A\&A{\rm,}\ }}

\def\mnras  {{MNRAS{\rm,}\ }}

\shortauthors{Lubin et al}
\shorttitle{X-ray Results on Optically-Selected Clusters}

\begin{document}

\title{The First Detailed X-ray Observations of High-Redshift,
Optically-Selected Clusters: XMM-Newton Results for Cl 1324+3011 at $z =
0.76$ and Cl 1604+4304 at $z = 0.90$}

\author{Lori M.~Lubin\altaffilmark{1}}
\altaffiltext{1}{Department of Physics, University of California, One 
Shields Avenue, Davis, CA 95616; lmlubin@ucdavis.edu}

\author{John S.~Mulchaey\altaffilmark{2}}
\altaffiltext{2}{The Observatories of the Carnegie Institution of Washington, 
813 Santa Barbara St., Pasadena, CA 91101; mulchaey@ociw.edu}

\author{Marc Postman\altaffilmark{3}}
\altaffiltext{3}{Space Telescope Science Institute, 3700 San Martin
Drive, Baltimore, MD 21218; postman@stsci.edu}

\begin{abstract}

We present the first detailed X-ray observations of optically-selected
clusters at high redshift. Two clusters, Cl 1324+3011 at $z = 0.76$
and Cl 1604+4304 at $z = 0.90$, were observed with XMM-Newton. The
optical center of each cluster is coincident with an extended X-ray
source whose emission is detected out to a radius of $\sim
0.5~h_{70}^{-1}$ Mpc. The emission from each cluster appears
reasonably circular, with some indication of asymmetries and more
complex morphologies. Similarly to other optically-selected clusters
at redshifts of $z \ga 0.4$, both clusters are modest X-ray emitters
with bolometric luminosities of only $L_x^{\rm bol} = 1.4-2.0
\times 10^{44}~h_{70}^{-2}$ erg s$^{-1}$. We measure gas temperatures
of $T = 2.88^{+0.71}_{-0.49}$ keV for Cl 1324+3011 and
$2.51^{+1.05}_{-0.69}$ keV for Cl 1604+4304. The X-ray properties of
both clusters are consistent with the high-redshift $L_x-T$ relation
measured from X-ray--selected samples at $z \ge 0.5$. However, based
on the local relations, their X-ray luminosities and temperatures are
low for their measured velocity dispersions ($\sigma$). The clusters
are cooler by a factor of 2--9 compared to the local $\sigma-T$
relation. We briefly discuss the possible explanations for these
results.

\end{abstract}


\section{Introduction} 

Clusters of galaxies provide a powerful probe of the nature of galaxy
formation and the origin of structure; thus, quantifying their
abundance and dynamical state is key to understanding the evolution of
galaxies and their environment.  Because X-ray luminosity ($L_x$)
scales as $({\rm density})^2 \times ({\rm temperature)}^{1/2}$,
cluster identification based on $L_{x}$ preferentially selects the
highest gas density regions; however, according to hierarchical
structure formation, we expect, at early times, to observe clusters
comprised of smaller ``proto-clusters'' with cooler gas (e.g., Frenk
et al.\ 1996). Some evidence for this may be the breakdown in the
X-ray--optical relations of moderate-to-high--redshift clusters which
are selected optically.  Specifically,
X-ray observations of all optically-selected clusters at $z \ga 0.4$
indicate that they are weak X-ray sources, regardless of their
measured richness or velocity dispersion, with luminosities of less
than a few $\times~10^{44}$ erg s$^{-1}$ (Castander et al.\ 1994;
Bower et al.\ 1994, 1997; Holden et al.\ 1997; Lubin, Oke \& Postman
2002). As a result, optically-selected clusters at $z \ga 0.4$ do not
obey the local relation between X-ray luminosity and velocity
dispersion. Their X-ray luminosities are low for their velocity
dispersions, indicating that optically-selected clusters at these
redshifts are underluminous (by up to a factor of $\sim 40$) compared
to their X-ray--selected counterparts (see Figures 5 of Bower et al.\
1997 and Lubin et al.\ 2002).

Very little is known about the specific causes of these differences as
only the X-ray--selected clusters at high redshift, and largely the
most luminous ($\ga 3 \times 10^{44}~h^{-2}_{70}$ erg s$^{-1}$) of
these, have been the subject of more detailed X-ray studies (e.g.,
Donahue et al.\ 1999; Gioia et al.\ 1999; Ebeling et al.\ 2001;
Stanford et al.\ 2001, 2002; Vikhlinin et al.\ 2002; Arnaud et al.\
2002; Jones et al.\ 2003; Valtchanov et al.\ 2003).
To address this issue, we have begun an XMM-Newton program to obtain
structural and spectral data on the most well-studied sample of
optically-selected clusters at $z \ga 0.6$, that of Oke, Postman \&
Lubin (1998). In this paper, we present the results from observations
of Cl 1324+3011 at $z = 0.76$ and Cl 1604+4304 at $z = 0.90$.  
We adopt $\Omega_m = 0.3$, $\Omega_\Lambda = 0.7$, and $H_0 =
70~h_{70}$ km s$^{-1}$ Mpc$^{-1}$.

\section{The XMM-Newton Observations} 

Cl 1324+3011 was observed by XMM-Newton on December 12-13, 2001 for a
total exposure of 39.6 ksec, while Cl 1604+4304 was observed on
February 9-11, 2002 for a total exposure of 43.2 ksec. All data
reduction and calibration was carried out using the XMM-Newton Science
Analysis System.  Light curves were created to identify times of high
background. Time intervals with count rates greater than 5 ct/s in the
MOS cameras and 10 ct/s in the PN camera were discarded. For Cl
1324+3011, the final exposure time is $\sim 32$ ksec. The Cl 1604+4304
observations suffered from major flaring activity resulting in total
usable exposure times of only $\sim$ 15 ksec for the PN and $\sim$ 20
ksec for the MOS cameras.

\ifsubmode
\else
\begin{figure}
\plotone{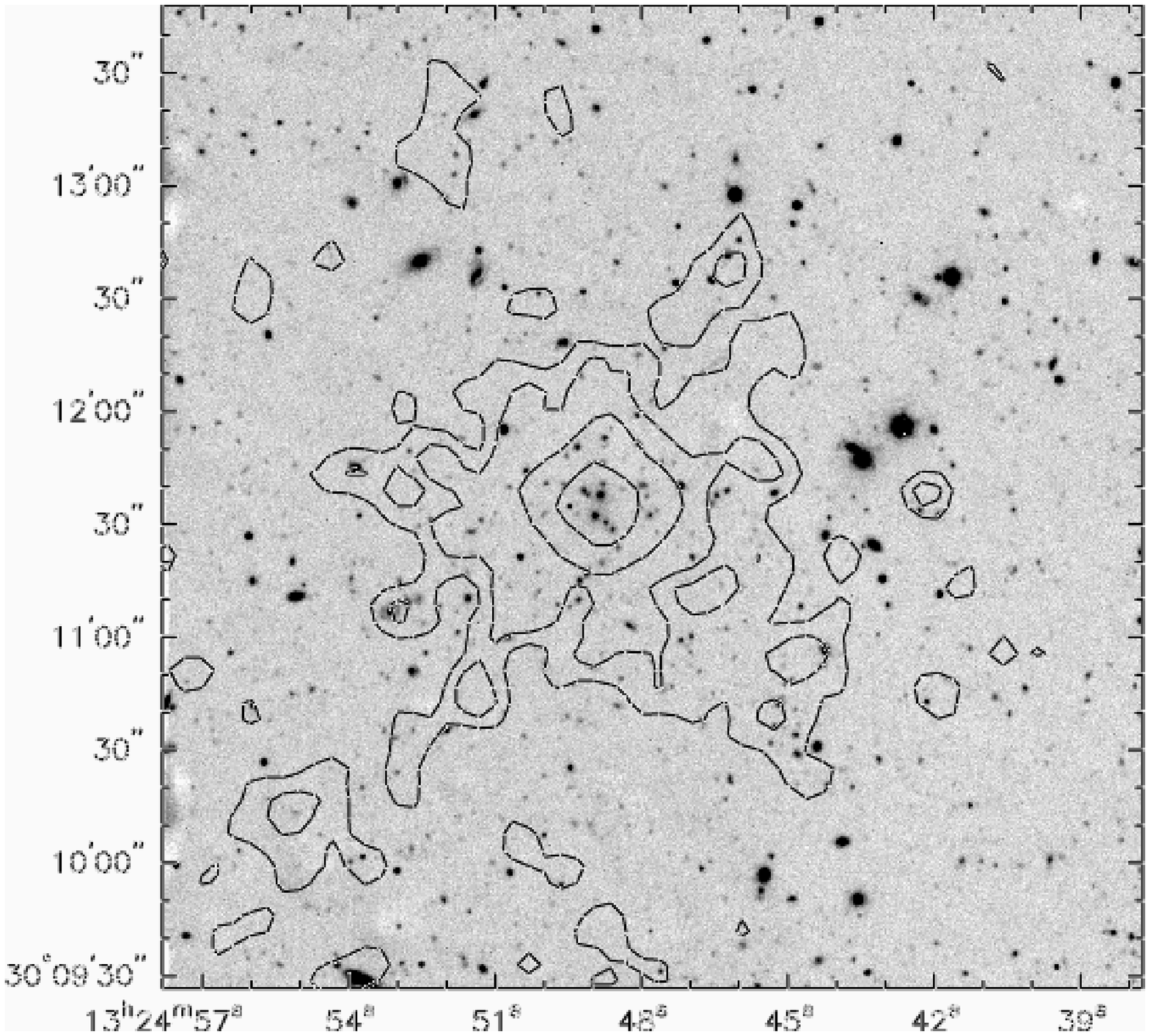}
\caption{The XMM-Newton contours of X-ray emission in Cl 1324+3011
overlaid on the Keck $R$-band image. Contours correspond to 5$\sigma$,
10$\sigma$, 20$\sigma$ and 40$\sigma$ above the background.  The X-ray
data have been smoothed with a Gaussian of width 7$''$. }
\label{13image}
\end{figure}
\fi

Total images in the 0.3--10 keV band were created by combining the
events from all three cameras. Both clusters are clearly detected as
extended sources.  To determine the extent of the emission, we
produced azimuthally-averaged surface brightness profiles for each
group with point sources masked out.  At a level corresponding to
3-sigma above the background, X-ray emission is detected out to a
radius of $\sim$ 100$''$ (514 $h^{-1}_{70}$ kpc) and $\sim$ 90$''$
(490 $h^{-1}_{70}$ kpc) for Cl 1324+3011 and Cl 1604+4304,
respectively.  The total number of source counts within these
apertures is $\sim$ 900 for Cl 1324+3011 and $\sim$ 500 for Cl
1604+4304.  Figures 1 and 2 show the XMM-Newton contours overlaid on
Keck $R$-band images. As can be seen from Figure 2, there is another
extended source to the southeast of Cl 1604+4304. Because there is no
redshift information on the galaxies coincident with this emission,
the nature of this source is unclear, and we have not included this
region in our analysis.  We fit the two-dimensional surface brightness
profile of each cluster using the fitting routine Sherpa, which is
part of the Chandra Interactive Analysis of Observations software
package.  The model was first convolved with the point spread
function.  The free parameters in the fit are the core radius ($r_c$),
$\beta$, and the background constant. For Cl 1324+3011, there were
enough counts to allow the ellipticity of the $\beta$ model to vary.
Both clusters are well-described by a $\beta$ model, and the best-fit
parameters (see Table 1) are consistent with values found for other
rich clusters (e.g., Mohr, Mathiesen \& Evrard 1999).

X-ray spectra were extracted for each camera using the apertures
described above. A local region was used for the background.  To
derive the spectral properties of the diffuse gas, we fit the spectra
with an absorbed MekaL model (Mewe, Gronenschild \& van den Oord 1985;
Kaastra \& Mewe 1993; Liedahl, Osterheld \& Goldstein 1995) using the
software package XSPEC.  In each case, we fixed the absorbing column
to the Galactic value given in Dickey \& Lockman (1990) and the gas
metallicity to 0.3 solar.  The results of the spectral fits are
summarized in Table 1, where all errors correspond to 1 sigma.

\ifsubmode
\else
\begin{figure}
\plotone{f2.ps}
\caption{The XMM-Newton contours of X-ray emission in Cl 1604+4304
 overlaid on the Keck $R$-band image. Contours correspond to
5$\sigma$, 10$\sigma$ and 20$\sigma$ above the background. The X-ray
data have been smoothed with a Gaussian of width 7$''$. }
\label{16image}
\end{figure}
\fi

The X-ray and optical centers of the clusters are well matched. The
optical center, defined as the mean position of all
spectroscoptically-confirmed cluster members (Postman, Lubin \& Oke
1998; Lubin et al.\ 2002), is within $8''$ and $11''$ of the center of
the X-ray contours for Cl 1324+3011 and Cl 1604+4304, respectively.
The X-ray contours of both clusters are reasonably circular, although
Cl 1604+4304 is slightly elongated in the north-south direction.  Both
clusters show indications of more complex morphologies similar to
X-ray--selected clusters at these redshifts (e.g., Gioia et al.\ 1999;
Stanford et al.\ 2001, 2002; Holden et al.\ 2002; Jones et al.\ 2003).

Both clusters are modest X-ray emitters with bolometric luminosities
of only $1.4$ and $2.0 \times 10^{44}~h^{-2}_{70}$ erg s$^{-1}$ for Cl
1324+3011 and Cl1604+4304, respectively (see also Castander et al.\
1994). Their measured temperatures are less than 3 keV (see Table 1),
consistent with the expectation from the local $L_x - T$ relation
(Mushotzky \& Scharf 1997; Markevitch 1998; Horner 2001). Modest
cooling in the central cores of both clusters is likely, with cooling
radii of $\sim 100$ kpc and cooling rates of $\sim 10$
M$_{\odot}$/yr. For Cl 1324+3011, we have sufficient counts to
re-analyze the data excluding the central cooling regions. The
resulting temperature is only 14\% larger than, and well within the
errors of, our original measurement, suggesting that the effect of
cooling is not large.

\section{Comparison to X-ray--Selected Clusters}


\ifsubmode
\else
  \begin{figure}
\plotone{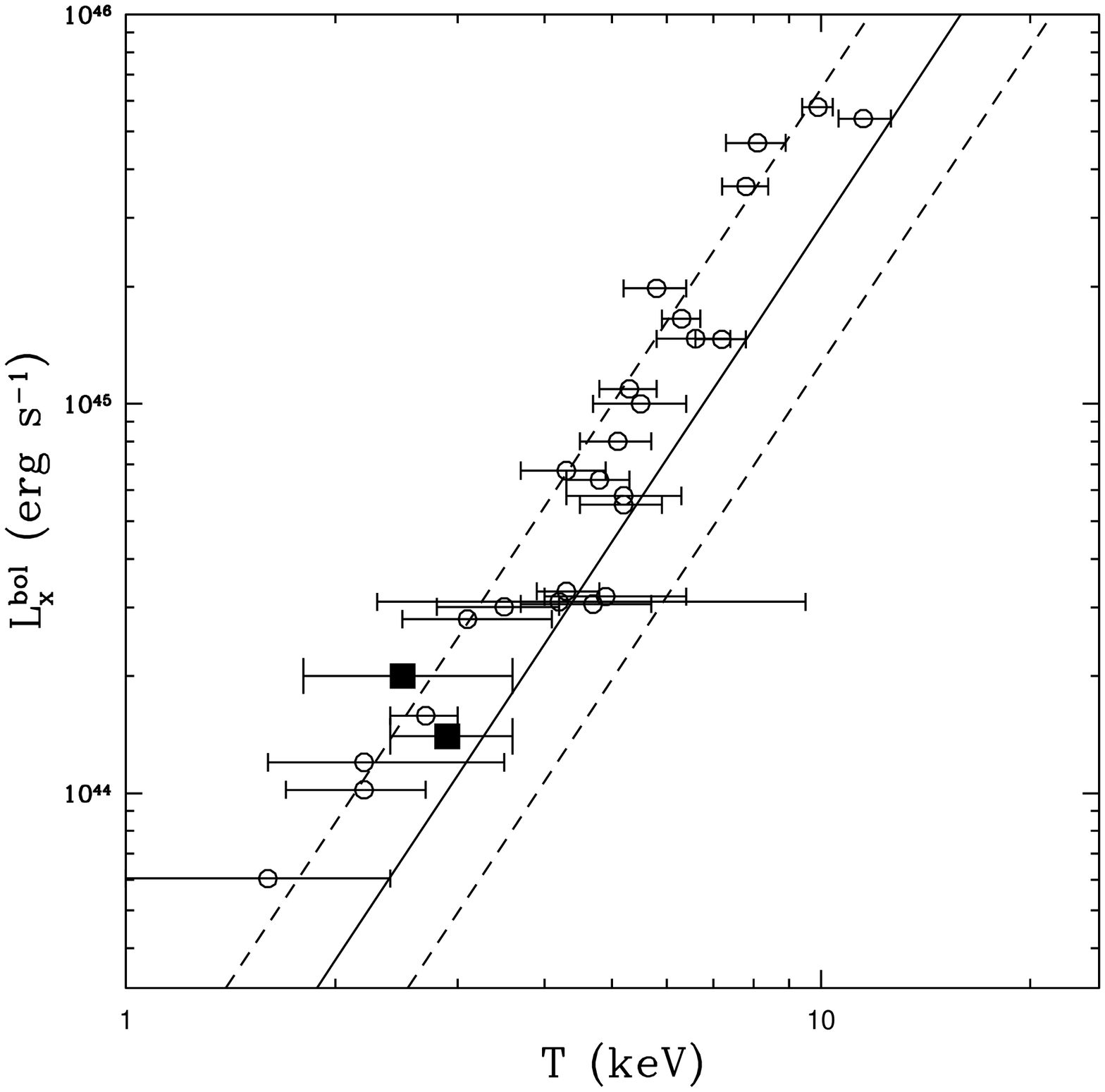}
\caption{The relation between temperature and bolometric luminosity 
for clusters at $z \ge 0.5$. Closed squares indicate the two
optically-selected clusters, Cl 1324+3011 and Cl 1324+3011. Open
circles indicate X-ray--selected clusters from Stanford et al.\
(2001), Arnaud et al.\ (2002), Hashimoto et al.\ (2002), Vikhlinin et
al.\ (2002), Jones et al.\ (2003), Maughan et al.\ (2003), and
Valtchanov et al. (2003). The solid line represents the best-fit,
least-squares line, and the dashed lines indicate the rms scatter
about the best-fit relation, as measured from the Horner (2001) sample
of clusters with $z < 0.5$ and $L_x^{\rm bol} > 10^{43}~h_{70}^{-2}$
erg s$^{-1}$.
}
\label{lx-t}
\end{figure}
\fi

\subsection{The $L_x - T$ Relation}

Figure 3 shows the relation between bolometric luminosity and
temperature for Cl 1324+3011 and Cl 1604+4304, as well as
X-ray--selected clusters at $z \ge 0.5$. The two optically-selected
clusters have X-ray properties which are consistent with those of the
X-ray--selected clusters, and they follow well the high-redshift
relation between luminosity and temperature. We observe no significant
differences, at least with these two systems, between the $L_x^{\rm
bol}-T$ relation for X-ray versus optically-selected clusters at these
redshifts.

We compare the high-redshift data to a large sample of clusters at
low-to-moderate redshifts observed by {\it ASCA} and uniformly
analyzed by Horner (2001). The sample contains 273 groups and
clusters, selected in both the optical and X-ray, which vary
significantly in galaxy and gas properties. Clusters with known
substructure and cooling flows are not excluded. In Figure 3, we plot
the best-fit $L_x^{\rm bol}-T$ relation measured from 233 clusters in
this sample, all of which have redshifts of $z < 0.5$ and luminosities
of $L_x^{\rm bol} > 10^{43}~h^{-2}_{70}$ erg s$^{-1}$.
Although Horner (2001) does not attempt to minimize the effect of
cooling flows on his measurements, the best-fit relation to these data
are consistent with previous measurements made from smaller samples
where the cooling regions have been excluded (e.g., Markevitch
1998). As noted previously, Figure 3 implies evolution in the
$L_x^{\rm bol}-T$ relation, with the high-redshift relation having a
larger luminosity for a fixed temperature (see also Fairley et al.\
2000; Holden et al.\ 2002; Novicki, Sorrig \& Henry 2002; Vihklinin et
al.\ 2002). However, the data points of the two optically-selected
clusters are still consistent with, and well within the measured rms
scatter about, the best-fit $L_x^{\rm bol}-T$ relation at low
redshift.

\subsection{The $\sigma - T$ Relation}

While we observe no differences between our optically-selected
clusters and their X--ray--selected counterparts when examining the
X-ray--X-ray relations, we do observe strong differences in the
X-ray--optical relations.
Using spectroscopy taken at Keck, both Cl 1324+3011 and Cl 1604+4304
have accurately measured velocity dispersions of $1016^{+126}_{-93}$
km s$^{-1}$ (47 redshifts) and $1226^{+245}_{-154}$ km s$^{-1}$ (22
redshifts), respectively. These velocity dispersions were measured
using a 3-sigma clipping technique on the full redshift distribution
of galaxies within a $2' \times 8'$ region centered on each
cluster. The average separation of the member galaxies from each
cluster center is $\sim 0.5~h^{-1}_{70}$ Mpc, the radius out to which
we detect the X-ray emission. Dispersions measured within smaller,
fixed apertures have much larger uncertainties due to the smaller
number of velocities used; however, they do agree, within the errors,
with the measurements made using the full redshift data (Postman et
al.\ 1998; Lubin et al.\ 2002). Based on the measured velocity
dispersions and the local $L^{\rm bol}_x-\sigma$ relation, the X-ray
luminosities of these clusters are low by a factor of 3--40 (Postman
et al.\ 1998; Postman, Lubin \& Oke 2001; Lubin et al.\ 2002).

With the XMM-Newton data, we can now measure the relation between
velocity dispersion and temperature for these clusters. This
comparison is more physically meaningful since both the velocity
dispersion of the galaxies and the temperature of the intracluster
medium provide a measure of the overall mass of the system. Although
the local $\sigma-T$ relation shows a large, non-statistical scatter,
the average relation is consistent with $\sigma \propto T^{1/2}$,
indicating that both the galaxies and the gas are isothermal and in
hydrostatic equilibrium within a common potential (e.g., Edge \&
Stewart 1991; Lubin \& Bahcall 1993).  Moderate-redshift clusters up
to $z \sim 0.5$ exhibit a similar $\sigma-T$ relation (Mushotzky \&
Scharf 1997).

\ifsubmode
\else
\begin{figure}
\plotone{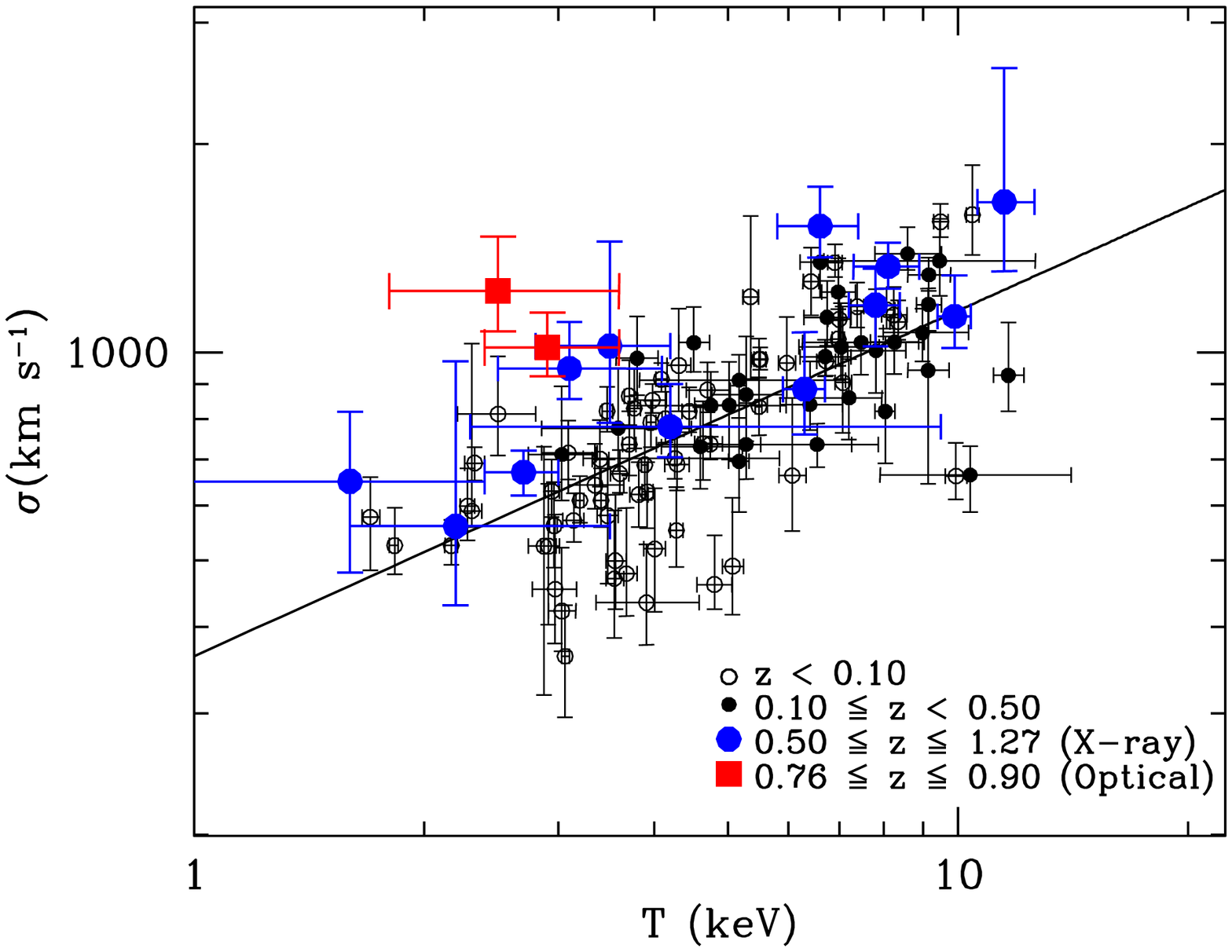}
\caption{The relation between temperature  and velocity dispersion 
for all clusters at redshifts of $z \le 1.27$. Small open and closed
circles indicate clusters with $L_x^{\rm bol} > 10^{43}~h_{70}^{-2}$
erg s$^{-1}$ at $z < 0.1$ and $0.1 \le z < 0.5$, respectively (Horner
2001). Red squares show the results for the two optically-selected
clusters, Cl 1324+3011 and Cl 1604+4304. Blue circles indicate
X-ray--selected clusters at $0.5 \le z \le 1.27$ (Borgani et al.\
1999; Donahue et al.\ 1999; Gioia et al.\ 1999; Tran et al.\ 1999;
Ebeling et al.\ 2001; Holden et al.\ 2001; Stanford et al.\
2001, 2002; Vikhlinin et al.\ 2002; Jones et al.\ 2003; Valtchanov et
al.\ 2003). The solid line shows the best-fit relation, given $\sigma
\propto T^{1/2}$, to the cluster data at $z < 0.5$.}
\label{sig-t}
\end{figure}
\fi
 
This correlation, and its significant scatter, is obvious in Figure 4
where we show the $\sigma-T$ relation for clusters at low-to-moderate
redshift taken from the Horner (2001) sample. For comparison, we plot
our two optically-selected clusters, as well as all X-ray--selected
clusters at $z \ge 0.5$ with measured velocity dispersions. The 11
high-redshift, X-ray--selected clusters are consistent, within the
errors, with the local $\sigma-T$ relation; however, there is an
indication of a systematic offset from this relation, with a
temperature that is higher, on average, by a factor of $\sim 1.4$ for
a given velocity dispersion. Because the errors on most of the
velocity dispersions, and some of the temperatures, are large, we
cannot ascertain this difference for certain.

Our optically-selected clusters, on the other hand, appear as clear
outlyers in this figure, with both clusters having a significantly
lower temperature than expected from their velocity dispersions. Based
on the best-fit relation to the data at $z < 0.5$, we estimate that Cl
1324+3011 is cooler by a factor of 2--4, while Cl 1604+4304 is cooler
by a factor of 3--9. However, if we include both the intrinsic scatter
in the low-redshift $\sigma-T$ data and the uncertainties in our
temperature and velocity dispersion measurements, these results may be
only a 4-sigma effect for Cl 1324+3011 and 5-sigma effect for Cl
1604+4304.  We clearly need accurate temperature measurements for a
larger sample of high-redshift, optically-selected clusters to
determine if they all systematically fall above the local $\sigma-T$
relation.

\section{Conclusions and Discussion} 

We have presented the first detailed spatial and spectral studies of
the intracluster medium in optically-selected clusters at high
redshift using observations from XMM-Newton. The two clusters, Cl
1324+3011 at $z = 0.76$ and Cl 1604+4304 at $z = 0.90$, have very
modest X-ray luminosities and are significantly underluminous for
their measured velocity dispersions when compared to the local
$L_x-\sigma$ relation. Each cluster has a correspondingly low
temperature which is consistent with both the low-redshift $L_x - T$
relation and the high-redshift relation measured previously from
purely X-ray--selected clusters at $z \ge 0.5$. This result suggests
that the intracluster medium behaves similarly in both optically and
X-ray--selected clusters.

The most obvious differences between the X-ray and optically-selected
clusters at high redshift occur in the X-ray--optical relations, such
as the $L_x-\sigma$ mentioned above.  We have now measured the
$\sigma-T$ relation for Cl 1324+3011 and Cl 1604+4304. Both clusters
are outlyers when compared to cluster data at $z < 0.5$. While
X-ray--selected clusters at $z \ge 0.5$ are more consistent with the
best-fit $\sigma \propto T^{1/2}$ relation, both optically-selected
clusters fall well above this relation, with temperatures that are
cooler by factors of 2--9.

The exact cause of these results is still unclear. The evolution in
the X-ray--optical relations may result from the fact that clusters at
these epochs are still in the process of forming. On the optical side,
we may measure an artificially high velocity dispersion because the
cluster is embedded in a filament oriented along the line-of-sight or
because there is a high rate of infall into the cluster environment
(see Bower et al.\ 1997). While no obvious substructure is observed in
either the angular or the redshift distribution of Cl 1324+3011 and Cl
1604+4304, it is not possible to determine whether these systems are
truly virialized using the current data (Postman et al.\ 1998; Lubin
et al.\ 2002).  However, we know that at least one of the clusters, Cl
1604+4304, is a member of a large scale structure (Lubin et al.\
2000), and we expect, based on semi-analytic models, that the infall
rate increases strongly with look-back time (Kauffmann 1995).

On the X-ray side, the existence of substructure and/or the merging of
subclumps can cause substantial changes in the properties of the
intracluster medium. Simulations of cluster formation between $z = 1$
and the present indicate a complex evolution in the emission-weighted
temperature and surface brightness maps. Initially, the gas is
distributed among many smaller, cooler subclumps, while, at
intermediate times, incomplete relaxation results in substantial
substructure in the system (e.g., Figures 2 \& 3 of Frenk et al.\
1996). For mergers of relatively equal-mass subclumps, the temperature
and luminosity of the gas component are significantly boosted, with
$T/L_x$ changing by several factors during the merger (Ricker
\& Sarazin 2001).
Because of the complexity of these physical processes, we require
considerably larger samples of well-studied clusters at $z \ga 0.5$,
including lower luminosity systems such as the optically-selected
clusters studied here, in order to characterize the effects of
dynamical evolution on the observed X-ray and optical properties.

\acknowledgements 

We would like to thank the anonymous referee, C.D.\ Fassnacht, and
J.B.\ Oke for their useful comments and essential contributions to
this paper.  Support for this work was provided by NASA through grant
NAG5-12373.

\ifsubmode
  \clearpage
\fi

\begin{deluxetable}{ccccccccc}
\tabletypesize{\footnotesize}
\tablecaption{Cluster X-ray Properties}
\tablewidth{0pt}
\tablehead{ & 
& \colhead{$T$} 
& \colhead{$L_x^{\rm 0.5-2.0~keV}$ \tablenotemark{(a)}} 
& \colhead{$L_x^{\rm bol}$ \tablenotemark{(a)}}
& \colhead{$r_c$ \tablenotemark{(b)}} & & & \\
\colhead{Name} 
& \colhead{$z$} 
& \colhead{(keV)} 
& \colhead{($\times 10^{44}$ erg s$^{-1}$)}
& \colhead{($\times 10^{44}$ erg s$^{-1}$)}
& \colhead{(kpc)} 
& \colhead{$\beta$ \tablenotemark{(b)}}
& \colhead{Ellipticity \tablenotemark{(b)}}}
\startdata
Cl 1324+3011 & 0.7565 & $2.88^{+0.71}_{-0.49}$ & 0.59 & 1.40 & $117 \pm 22$& $0.78^{+0.19}_{-0.12}$ & $0.21^{+0.23}_{-0.21}$ \\
Cl 1604+4304 & 0.8967 & $2.51^{+1.05}_{-0.69}$ & 0.86 & 2.01 & $144 \pm 7\phantom{0}$ & $0.77^{+0.04}_{-0.03}$ & \nodata \\
\enddata
\tablenotetext{(a)}{Total X-ray luminosity in the 0.5--2.0 keV band and the 
bolometric luminosity are measured within a radius of 514 and 490
$h^{-1}_{70}$ kpc for Cl 1324+3011 and Cl 1604+4304, respectively.}
\tablenotetext{(b)}{The spatial distribution of emission  
is fit to an elliptical, two-dimensional $\beta$ model for Cl
1324+3011 and to only a circular, two-dimensional $\beta$ model for Cl
1604+4304 due to the smaller number of counts.}
\end{deluxetable}

\ifsubmode
\clearpage
\begin{figure}
\plotone{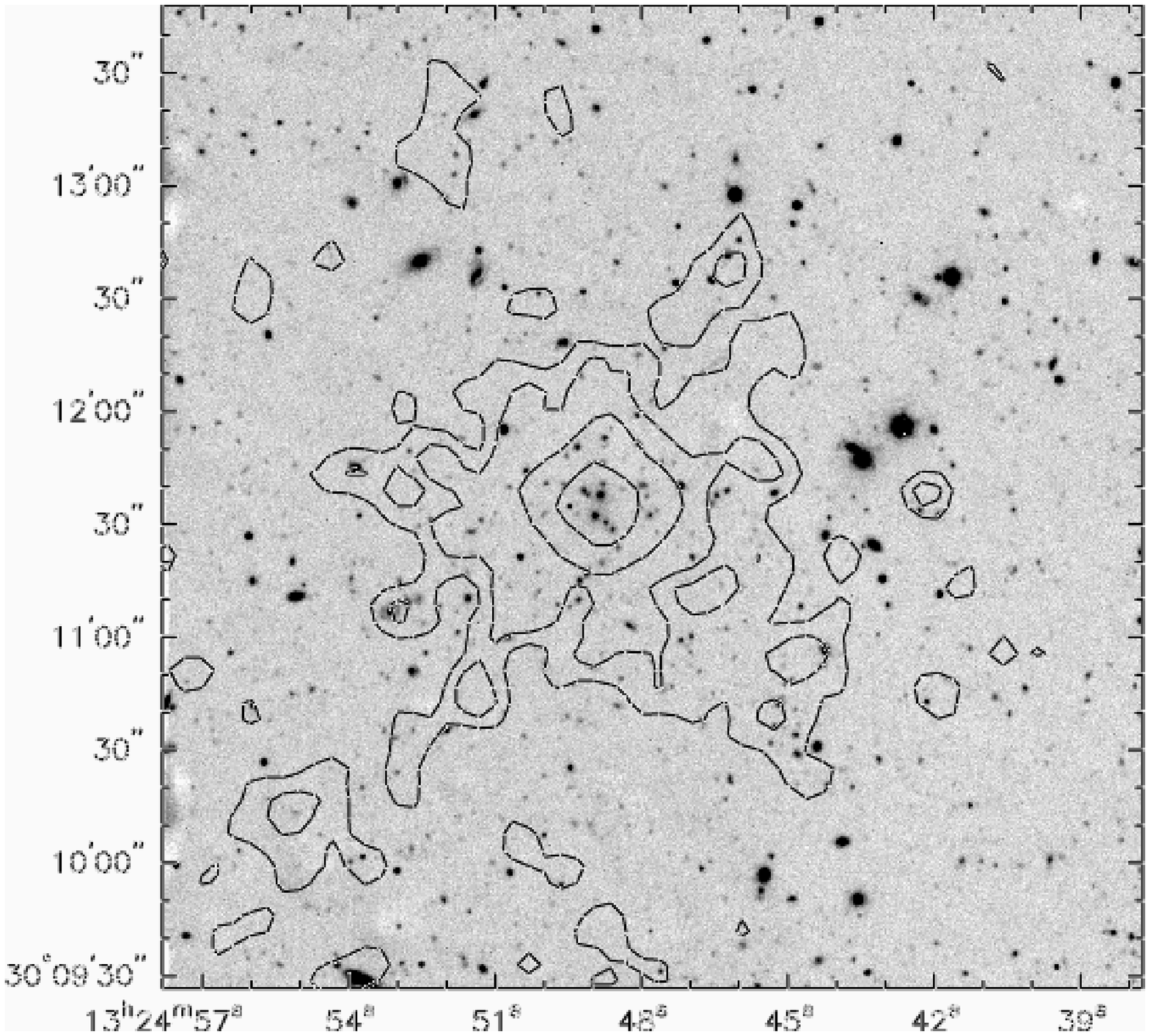}
\caption{The XMM-Newton contours of X-ray emission in Cl 1324+3011
overlaid on the Keck $R$-band image. Contours correspond to 5$\sigma$,
10$\sigma$, 20$\sigma$ and 40$\sigma$ above the background.  The X-ray
data have been smoothed with a Gaussian of width 7$''$. }
\label{13image}
\end{figure} 
\fi

\ifsubmode
\clearpage
\begin{figure}
\plotone{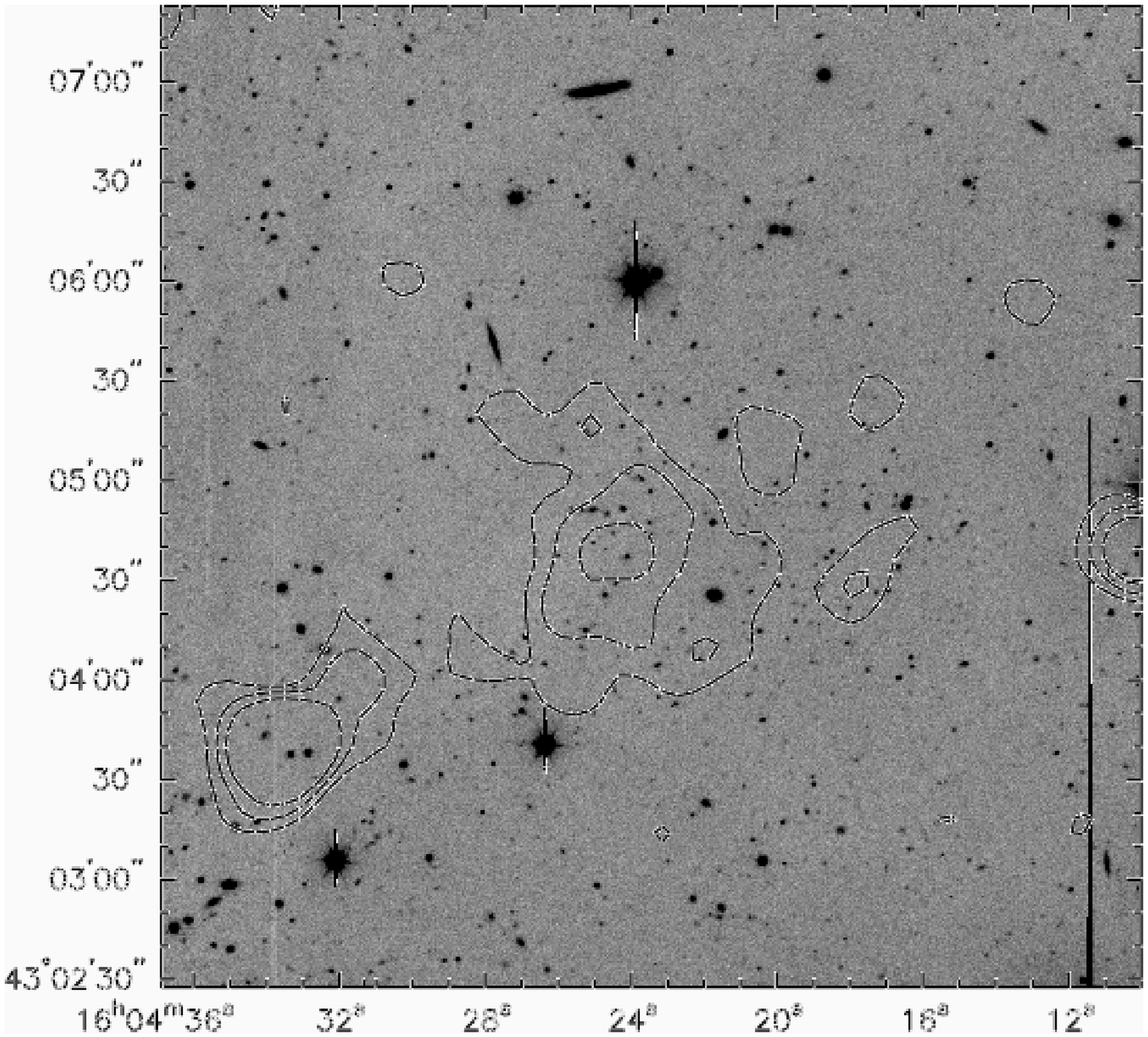}
\caption{The XMM-Newton contours of X-ray emission in Cl 1604+4304
 overlaid on the Keck $R$-band image. Contours correspond to
5$\sigma$, 10$\sigma$ and 20$\sigma$ above the background. The X-ray
data have been smoothed with a Gaussian of width 7$''$. }
\label{16image}
\end{figure}
\fi

\ifsubmode
\clearpage
\begin{figure}
\plotone{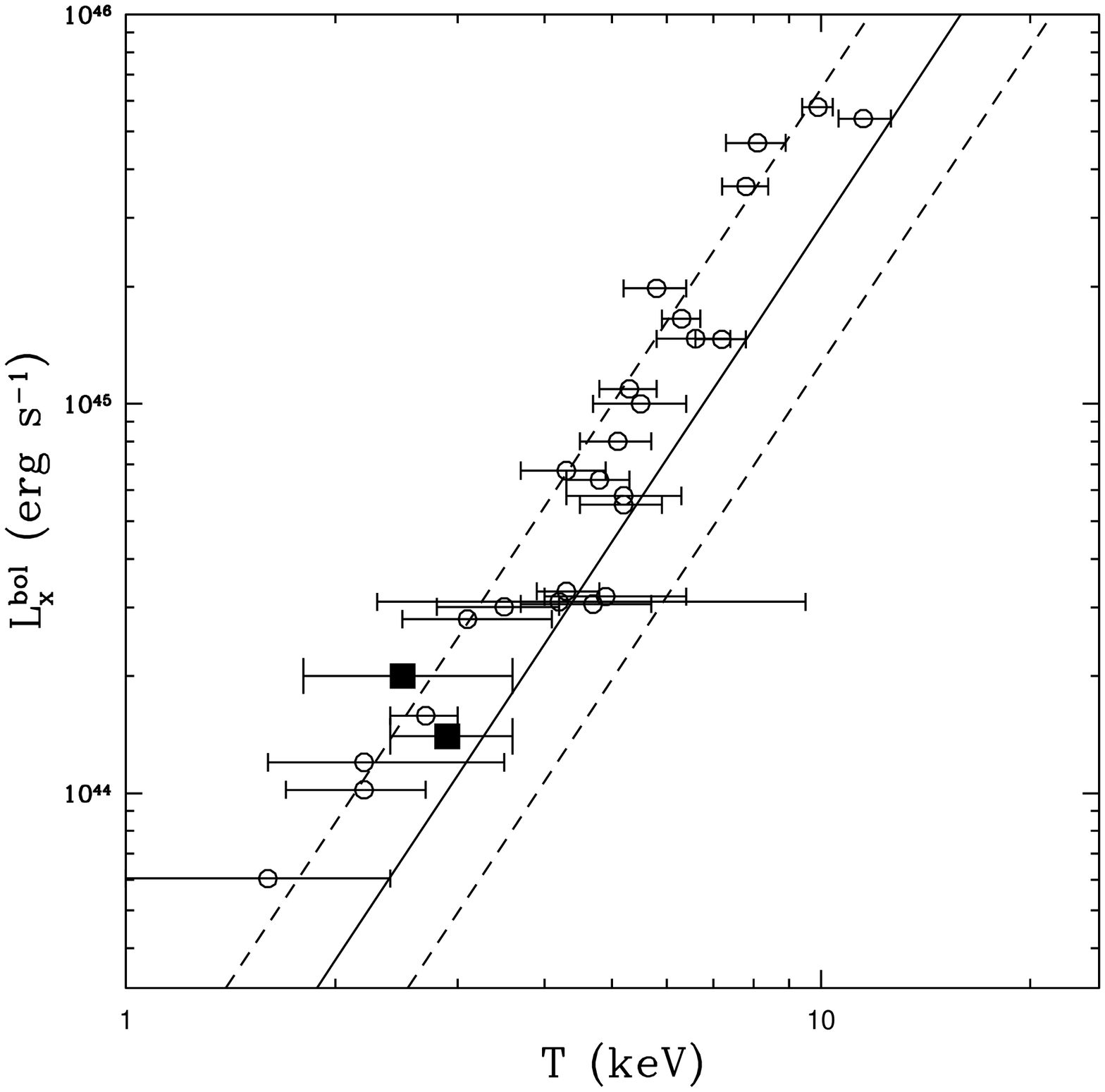}
\caption{The relation between temperature and bolometric luminosity 
for clusters at $z \ge 0.5$. Closed squares indicate the two
optically-selected clusters, Cl 1324+3011 and Cl 1324+3011. Open
circles indicate X-ray--selected clusters from Stanford et al.\
(2001), Arnaud et al.\ (2002), Hashimoto et al.\ (2002), Vikhlinin et
al.\ (2002), Jones et al.\ (2003), Maughan et al.\ (2003), and
Valtchanov et al. (2003). The solid line represents the best-fit,
least-squares line, and the dashed lines indicate the rms scatter
about the best-fit relation, as measured from the Horner (2001) sample
of clusters with $z < 0.5$ and $L_x^{\rm bol} > 10^{43}~h_{70}^{-2}$
erg s$^{-1}$.
}
\label{lx-t}
\end{figure}
\fi

\ifsubmode
\clearpage
\begin{figure}
\plotone{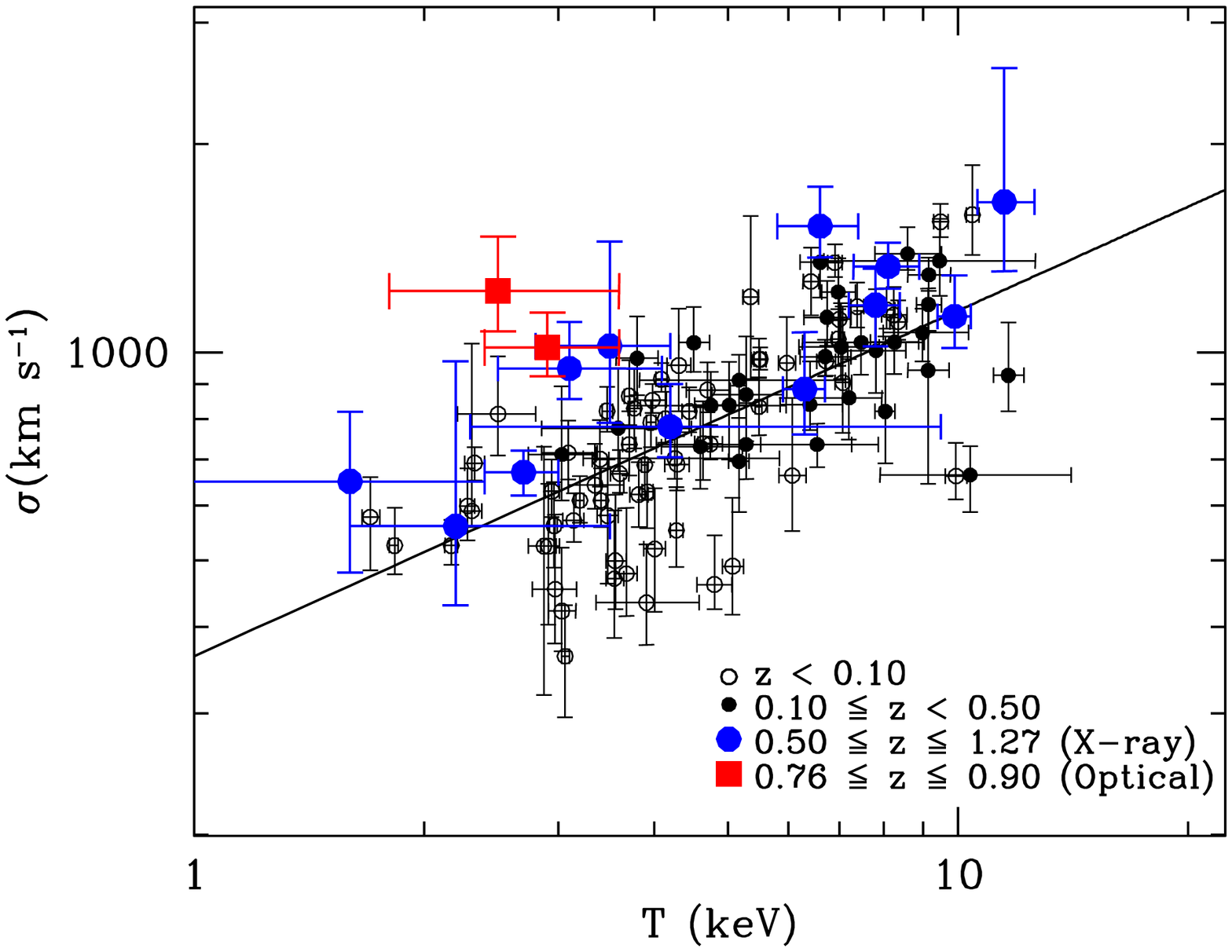}
\caption{The relation between temperature  and velocity dispersion 
for all clusters at redshifts of $z \le 1.27$. Small open and closed
circles indicate clusters with $L_x^{\rm bol} > 10^{43}~h_{70}^{-2}$
erg s$^{-1}$ at $z < 0.1$ and $0.1 \le z < 0.5$, respectively (Horner
2001). Red squares show the results for the two optically-selected
clusters, Cl 1324+3011 and Cl 1604+4304. Blue circles indicate
X-ray--selected clusters at $0.5 \le z \le 1.27$ (Borgani et al.\
1999; Donahue et al.\ 1999; Gioia et al.\ 1999; Tran et al.\ 1999;
Ebeling et al.\ 2001; Holden et al.\ 2001; Stanford et al.\ 2001,
2002; Vikhlinin et al.\ 2002; Jones et al.\ 2003; Valtchanov et al.\
2003). The solid line shows the best-fit relation, given $\sigma
\propto T^{1/2}$, to the cluster data at $z < 0.5$.}  \label{sig-t}
\end{figure} 
\fi 

\end{document}